# Transient reconfigurable subangstrom-precise photonic circuits at the optical fiber surface

A. Dmitriev[(a)*], N. Toropov[(b)*], and M. Sumetsky[(a)*]

[(a)]Aston Institute of Photonic Technologies, Aston University, Birmingham B4 7ET, UK
[(b)]ITMO University, St.Petersburg, 197101, Russia
[*]a.dmitriev@aston.ac.uk; [*]nikita.a.toropov@gmail.com; [*]m.sumetsky@aston.ac.uk

Transient fully reconfigurable photonic circuits can be introduced at the optical fiber surface with subangstrom precision. A building block of these circuits – a 0.7Å-precise nano-bottle resonator – is experimentally created by local heating, translated, and annihilated.

## I. INTRODUCTION

Fabrication of microscopic optical processors, which can eventually compete with electronic processors in speed and power consumption, challenged researchers for decades [1, 2]. Around the year 2000, the slow light devices, which are based on circulation of light in coupled ring resonators and oscillation in photonic crystals, were suggested as a promising solution [3, 4]. These devices can potentially serve as miniature optical delay lines and buffers – the building blocks of the future optical and microwave signal processors, optical sensors and quantum computers.

The major obstacles to the realistic applications of slow light devices today are insufficient fabrication precision and attenuation of light in photonic circuits. In spite of remarkable progress (several nanometer precision has been achieved in silicon photonics [5, 6]), the precision and insertion loss of microphotonic circuits are still far beyond those required by applications [7, 8].

Recently, a special class of photonic circuits with dramatically improved fabrication precision and performance has been demonstrated based on the Surface Nanoscale Axial Photonics (SNAP) platform [9-11]. The SNAP circuits are fabricated at the extremely smooth and low-loss surface of an optical fiber. It has been shown that the exceptionally small nanoscale deformation of the effective fiber radius is sufficient to fully control slow whispering gallery mode (WGM) propagation along the fiber surface. The experimentally demonstrated ultralow-loss SNAP structures and devices, such as coupled ring resonators [10] and bottle resonator delay lines [11], are fabricated with the subangstrom precision approaching the practical requirements.

It is critical to develop the SNAP platform to create switchable, tunable, and, ideally, fully reconfigurable photonic circuits. This paper presents an imperative demonstration supporting the feasibility of these wide-reaching objectives. We show that a SNAP photonic circuit, which is introduced temporarily, can maintain its shape with subangstrom precision for the predetermined period of time and be completely annihilated under request. In particular, we experimentally demonstrate a building block of a SNAP photonic circuit, a nano-bottle resonator, which is introduced by local heating of the fiber with a focused $CO_2$ laser beam and translated along the fiber, continuously keeping its shape unchanged with the excellent precision of 0.7 Å. Finally, we investigate the millisecond-scale relaxation process by switching the heating on and off. The experimental results obtained are in very good agreement with theory.

## II. THEORY: THERMALLY INTRODUCED TRANSIENT SNAP PHOTONIC CIRCUITS

The temporary nanoscale variation of the effective fiber radius can be introduces by heating with a focused $CO_2$ laser beam (as in the experiment of Section III) as well as with electric current heating using metal nanowires and elements adjacent to fiber. For the conditions of our experiment, the temperature variation $T(x,t)$ along the fiber axis $x$ is described by a 1D heat equation $\rho c_p T_t - k T_{xx} + 2h r_0^{-1} T = Q(x,t)$, where $Q(x,t)$ is the laser beam power absorbed by the fiber, $r_0 = 19$ µm is the fiber radius, and other parameters are the silica fiber density $\rho = 2.2 \times 10^3$ kg/m$^3$, specific heat capacity $c_p = 703$ J/(kg·K), thermal conductivity $k = 1.38$ W/(m·K), and heat transfer coefficient $h \approx 350$ W/(m$^2$·K). Fig. 1(a) shows the calculated effective fiber radius variation $\Delta r(x) = (r_0/n)(dn/dT) T(x)$, where the silica refractive index $n = 1.44$ and $dn/dT = 1.2 \times 10^5$ K$^{-1}$. in the presence of a stationary Gaussian laser beam with the FWHM equal to 106 µm corresponding to the conditions of our experiment described in Sec. III. Figs. 1(b) and (c) show the evolution of this temperature profile in the process of heating and cooling, respectively. For comparison, Fig. 1(d), (e), and (f) show the results of similar calculations for the case of five Gaussian beams separated by 250 µm.

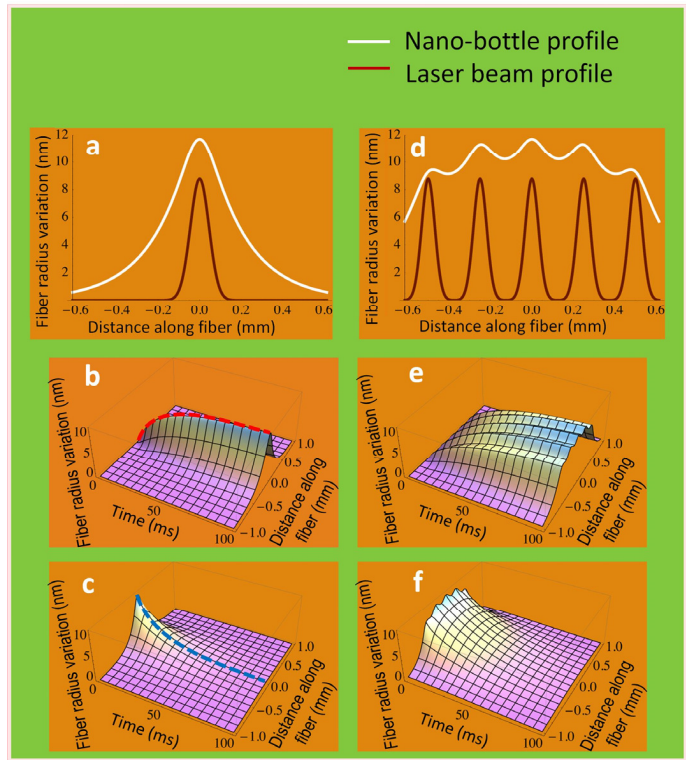

Fig. 1. Theory. (a) – The input power distribution along the fiber axis $x$ (brown curve) and corresponding nano-bottle radius variation $\Delta r(x)$ (white curve) introduced by stationary heating with a Gaussian laser beam. (b) – Evolution of radius variation shown in (a) in the course of initial heating. Dashed red line indicates the evolution of nano-bottle center radius variation. (c) – Evolution of radius variation shown in (a) after switching the laser beam off. Dashed blue line indicates the evolution of nano-bottle center radius variation. (d), (e) and (f) – Plots similar to those shown (a), (b), and (c) calculated for the fiber radius variation introduced by five 250 μm spaced Gaussian laser beams.

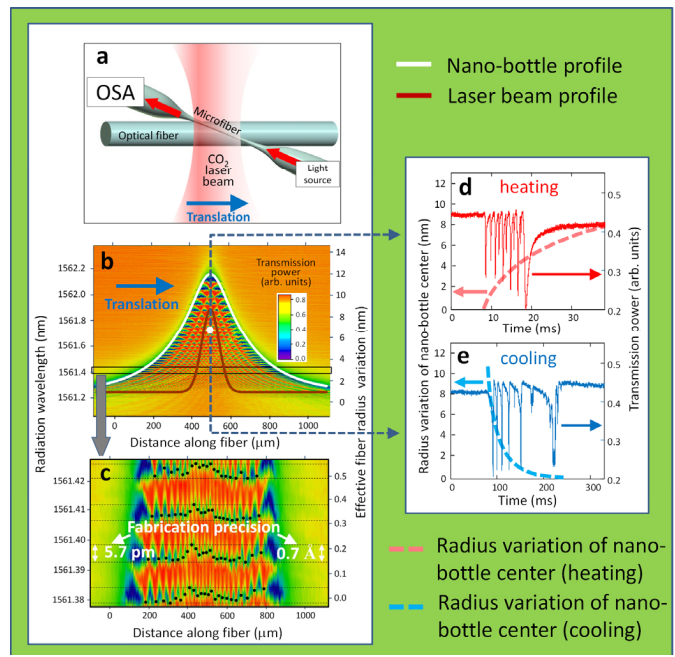

Fig. 2. Experiment. (a) – An optical fiber segment locally heated by a focused $CO_2$ laser beam which is translated along the fiber axis. (b) – The surface plot of the spectrum of a nano-bottle resonator measured in the process of the laser beam translation. Bold white and brown curves are the nano-bottle and input beam profiles theoretically calculated in Fig. 1(a). (c) – Magnified section of the surface plot outlined in (b) tilted by 0.16 nm/mm, to compensate for the linear fiber nonuniformity and translation misalignment. Black circles are the positions of high Q-factor spectral dips of the bottle resonator, which determine the fabrication precision 0.7 Å in effective fiber radius variation. (d) and (e) – Heating and cooling relaxation passages of resonant spectra measured at wavelength and position indicated by a white central point in Fig. 2(b) compared to the relaxation of radius variation of the nano-bottle center calculated in Fig. 1(b) and (c).

## III. EXPERIMENT: CREATION, TRANSLATION, AND ANNIHILATION OF A NANO-BOTTLE RESONATOR

Our experimental setup is sketched in Fig. 2(a). An optical fiber segment with radius $r_0 = 19$ μm is locally heated by a focused $CO_2$ laser beam, which leads to a nanoscale variation of the effective fiber radius $\Delta r(x)$ and the formation of a SNAP bottle resonator. WGMs are launched into the resonator with a transversely oriented biconical microfiber taper touching the fiber at a fixed position. Next, the nano-bottle is translated along the fiber axis following the continuously moving laser beam, while the WGM spectra are periodically detected by a Luna OSA after each 4.7 μm interval. The surface plot of the measured mode spectra is shown in Fig. 2(b). From this figure, the height of the travelling nano-bottle is approximately 12 nm (corresponding to the maximum temperature variation of 74 K), while the axial FWHMs of this resonator and its fundamental mode are 390 μm and 45 μm, respectively. The measured nano-bottle radius profile $\Delta r(x)$ (white curve in Fig. 2(b)) is in excellent agreement with theory (white curve in Fig. 1(a)). In the regions of weak coupling to the microfiber, the wavelengths of high Q-factor modes of an ideally stable resonator are independent of the measurement positions. Hence, the deviations of these resonances from horizontal spectral lines in Fig. 2(c) determine our fabrication and translation precision which is found equal to 0.7 Å in effective fiber radius variation. Finally, Fig. 2(d) and (e) show the relaxation passages of resonant peaks through the fixed wavelength determined by the white central point in Fig. 2(b) measured with a Tektronix oscilloscope. Both heating (Fig. 2(d)) and cooling (Fig. 2(e)) passages are in a good agreement with the calculated heating and cooling evolution of the nano-bottle center radius variation shown in Fig. 1(b) and (c). Following this demonstration, the introduction of more complex transient photonics circuits (e.g., that shown in Fig. 1(d)) becomes straightforward by the specially distributed input laser beam produced, e.g., with an acousto-optic modulator.

## IV. SUMMARY AND DISCUSSION

In summary, transient and fully reconfigurable photonic circuits can be introduced at the bare surface of an optical fiber with unprecedented subangstrom precision. This result paves the way to the creation of several important building blocks, structures, and devices for the future high-performance optical signal processors and sensors, including quantum processors. In particular, this result supports the feasibility of highly precise fully configurable ultralow-loss SNAP structures based on strongly nonlinear fibers, using e.g., silicon-core silica fibers which enable the required nanosecond-scale fast and nanometer-scale fiber radius variation tunability [12]. It has been suggested recently that a subangstrom precise nanosecond-fast reconfigurable nano-bottle resonator can solve the long standing problem of creating a practical microscopic optical buffer, the key element of the future optical signal processing devices [13, 14]. The ultraprecise millisecond-scale reconfigurability demonstrated in this paper for the thermally introduced nano-bottle supports the feasibility of such buffer. It is expected that it can be created at the surface of highly nonlinear silicon-core and more complex multimaterial fibers which are reconfigurable at nanosecond time scale with subangstrom precision.